**On conditions of negativity of friction resistance for non-stationary modes of blood flow and possible mechanism of affecting of environmental factors on energy effectiveness of cardio-vascular system functioning**

S.G. Chefranov

**Abstract**

It is shown that initiated by action of molecular viscosity impulse flow, directed usually from the moving fluid to limiting it solid surface, can, under certain conditions, turn to zero and get negative values in the case of non-stationary flow caused by alternating in time longitudinal (along the pipe axis) pressure gradient. It is noted that this non-equilibrium mechanism of negative friction resistance in the similar case of pulsating blood flow in the blood vessels, in addition to the stable to turbulent disturbances swirled blood flow structure providing, can also constitute hydro-mechanical basis of the observed but not explained yet paradoxically high energy effectiveness of the normal functioning of the cardio-vascular system (CVS). We consider respective mechanism of affecting on the stability of the normal work of CVS by environmental variable factors using shifting of hydro-dynamic mode with negative resistance realization range boundaries and variation of linear hydro-dynamic instability leading to the structurally stable swirled blood flow organization.

### 1. Introduction

Investigations of non-stationary flow of viscous incompressible fluid in the pipes with elastic and rigid walls are widely used for modeling of pulse blood flow in arteries [1-3]. The problem of observed mysteriously high energy effectiveness of CVS functioning is not yet solved. It can't be explained based on usual assumptions on rapid increase of resistance to the blood flow with successive branching of vessels and their narrowing [4]. Actually, according to the Poiseuille law, viscous flow friction on walls of blood vessels must grow as the fourth power of the value of relative decrease of the vessel's radius [4,5]. Hence, to pump the blood through the CVS (where the vessels diameter ranges from 10 mm to $10^{-3}$ mm) it would be necessary pressure difference dozens times greater than that is character for CVS artery-vein pressure difference (about 100 mm mercury column).

However, from the other side, Poisuille law itself is applicable to already stationary flows only, to which in the general case, non-stationary mode of real blood flow in CVS does not belong. It is also known that even for non-stationary modes in the pipes with rigid wall observed value of volumetric rate of flow can significantly differ from pre-calculated values and rate of its change in time grows with the decrease of the used in experiments pipe diameter [6].

J. Womersley (see [3] and referenced there his previous works) was one of the first who theoretically considered peculiarities character for very non-stationary modes in relation with modeling of artery blood flow in CVS. He introduced a dimensionless parameter $\alpha = R\sqrt{\lambda/\nu}$ (further – Womersley number), value of which characterizes extent of deviation of the viscous fluid flow under consideration from a stationary mode. Here R is the radius of a pipe along which the fluid with kinematical viscous coefficient $\nu$ flows, and value $\lambda^{-1}$ is defined by the character time scale of variability of longitudinal (along the pipe axis) pressure gradient providing existence of the fluid flow in the pipe. In particular, for the case of periodic variations in time of the longitudinal pressure gradient this time scale corresponds to the period of such variability [3]. As noted in [3], the parameter $\alpha$ is a modification of existed before this only one dimensionless parameter being a ratio of the width of the blood vessel wall to its radius. Also it is set in [3] that for some values of the parameter $\alpha$, relationships between the pressure gradient and stream velocity for elastic and rigid pipe walls are practically the same. It is considered that the Womersley number shows how much velocity profile for laminar flow in a long pipe differs from the Poiseuille's one when the fluid is subjected to variable in time longitudinal pressure gradient [1]. In the CVS values of $\alpha$ (calculated for the character heart beat rate) vary in the wide range. So, in aorta, $\alpha$ can be more than 10, whereas in capillaries it is about $10^{-3}$ [1].

In [3], however, there were not considered the modes of the pressure gradient change character for relatively short time intervals lying inside a heart cycle when only monotonic change (in the most part of the cycle it is monotonic decreasing) of this value takes place.

In the present work, it is made in particular in assumption of exponential monotonic pressure gradient decreasing with time that allowed defining explicit analytical conditions of existence of the flow modes with negative value of the viscous friction force for a flow with respect to the pipe walls with circular cross-cut of radius R. Womersley numbers are estimated for which conditions of linear instability of non-stationary Hagen-Poiseuille flow are met and it is possible emergence of the resulting spiral flow relatively more stable with respect to turbulent disturbances than a laminar non-swirled mode of the fluid flow in the pipe. Applications of the obtained conclusions to explaining of the observed paradoxically high energy effectiveness of CVS normal work are considered. Possible hydro-mechanical mechanism explaining impact of the varying environmental factors on emergence of weather induced pathologies level of which depends on the individual features of stable CVS work and other integrative organism systems in the whole.

## 2. Exact solution of non-stationary hydro-dynamic equations for a circular pipe

Let us consider the flow of viscous incompressible fluid in stationary straight pipe of the unbounded length having circular cross-section (constant for the whole pipe) with radius R. We assume that longitudinal along the pipe axis pressure gradient causing the fluid flow is as follows:

$$\frac{\partial p}{\partial z} / \rho_0 = -a_0 \exp(-\lambda t) = -a(t), \qquad (1)$$

Where in cylindrical reference frame (z, r, $\varphi$), coordinate z is directed along the pipe axis, and $\rho_0$ is the constant fluid density. We consider flow in the pipe caused by (1) when it is possible to count values of radial and tangent components of the velocity field in the pipe equal to zero, i.e. $V_\varphi = V_r = 0$.

For the longitudinal component of the velocity field $V_z$ taking into account (1), we ga=have the following expression for evolutionary equation followed from nonlinear non-stationary three-dimensional Navie-Stokes equations [5] :

$$\frac{\partial V_z}{\partial t} = a(t) + \frac{\nu}{r}\frac{\partial}{\partial r} r \frac{\partial}{\partial r} V_z, \qquad (2)$$

where the function $a(t)$ is defined in (1).

It is not difficult to show that an exact solution of the equation (2), meeting natural boundary condition of non-slipping (when on the solid boundary due to the action of viscous forces fluid flow velocity turns to zero, i.e. $V_z = 0$ for r=R) has the following form:

$$V_z = -\frac{a_0 R^2}{\alpha^2 \nu} \exp(-\alpha^2 \nu t / R^2) \left[ 1 - \frac{J_0(r\alpha/R)}{J_0(\alpha)} \right], \tag{3}$$

where $a_0 > 0, \lambda > 0$, $J_0$ is the Bessel function of the zero order, and the value of $\alpha = R\sqrt{\lambda/\nu}$ corresponds to the defined above Womersley number [3]. Contrary to [3], here value of $\lambda$ is real (i.e. it has zero imaginary part), whereas in [3] it is considered as purely complex with zero real part. In the rest, solution (3) coincides with formula (12) of the work [3] in the variant corresponding to the problem of finding an expression for a longitudinal velocity field component via longitudinal pressure gradient. Note that the specified variant is a particular case of the formula (10) from [3], taking into account also elastic features of the pipe wall when modeling artery blood flow.

Using (3) and definition of the tangent friction force per boundary surface square unit, $\sigma = \rho_0 \nu \left( \frac{\partial V_z}{\partial r} \right)_{r=R}$ [5] (see also in [5] the formula (15.17) for the component $\sigma_{zr} = \sigma$ of the viscous stress tensor), we get an exact expression:

$$\sigma = -\rho_0 R a_0 \exp(-\lambda t) \frac{J_1(\alpha)}{\alpha J_0(\alpha)}, \tag{4}$$

where $J_1$ is the Bessel function of the first order and the sign of $\sigma$ characterizes the direction of the impulse flow. So, for the negative values $\sigma < 0$, the impulse flow is directed as usually from the flowing fluid to the pipe wall causing the flow deceleration due to the positive force of the friction force with respect to the pipe wall. Conversely, from (4), it follows that for the values of $\alpha > 0$, for which the Bessel function of the first order turns to zero and then gets negative values, friction force also turns to zero and gets negative values when $\sigma > 0$ in (4). In particular, in the limit of large values $\alpha \gg 1$ in (4) the function $\frac{J_1(\alpha)}{J_0(\alpha)} \to tg\alpha$. In this limit, positivity $\sigma > 0$ in (4) takes place under condition of meeting the following inequalities (following from the condition $tg\alpha < 0$):

$$\pi/2 + \pi n < \alpha < \pi + \pi n, \tag{5}$$

where n is an integer value and in (5) it must be n>>1. In the general case, for $\alpha$, instead of (5), the following restrictions defined by the conditions of negativity of $J_1(\alpha)/J_0(\alpha) < 0$ are obtained:

$$\gamma_{0,n} < \alpha < \gamma_{1,n}, n = 1, 2, \ldots \tag{6}$$

where for any integer n by definition $J_1(\gamma_{1,n}) = 0; J_0(\gamma_{0,n}) = 0$ (in particular, $\gamma_{0,1} \approx 2.4; \gamma_{1,1} \approx 3.8$ и т.д.). Thus, scale invariance of the condition (6) takes place related with the possibility of its realization for finite and arbitrary large values of the Womersley number (the latter when selecting sufficiently large integer value of n in (6)).

When as in [3], value $\lambda = i\omega$ in (1)-(4), already in the general there are no such simple conditions of negativity of the friction force as (5), (6). However, in the limit of large values of the Womersley number (in the form just considered in [3]) $\alpha_0 = R\sqrt{\omega/\nu} \gg 1$ it is already not difficult to get instead of (5) the following condition (when replacing in (4) $\lambda$ by $i\omega$)

$$-\pi/4 + \pi(2k-1) < \omega t < -\pi/4 + \pi 2k, k \gg 1, \tag{7}$$

where k is an integer. Inequalities (7) define finite intervals of time corresponding to the negative friction force per the pipe square unit. Condition (7), contrary to (5),(6), is already independent from the pipe radius. It is obtained from the inequality

$$\sigma \square -\rho_0 a_0 \sqrt{\frac{\nu}{2\omega}}(\cos \omega t + \sin \omega t)(1+O(e^{-R\sqrt{2\omega/\nu}})) > 0 \qquad (8)$$

The condition (7) shows possibility of realization of energy effective mode with negative friction force (and impulse flow directed from the pipe wall to the stream) only for some time intervals but not for all time instances as in (5) и (6). However, such a mode with oscillating pressure gradient is more energy effective than stationary Poiseuille mode because on average for the period of the pressure gradient varying value of (8) turns to zero and losses on the friction have exponentially small value corresponding to the terms neglected in (8).

Note that the obtained conclusions about possibility of the sign change of the impulse flow on the pipe boundary under conditions (5) - (7) are not reflected on the estimate of the value of corresponding energy flow value of which on the boundary for the flow (3) in any case is zero.

Consider also an estimate of the value of hydro resistance that generalizes Poiseuille law for the case of stationary flow (3). Rate of the fluid in the pipe is defined by the volumetric velocity $Q = 2\pi \int_0^R dr r V_z$, which for the flow (3) has the following form:

$$Q = -\frac{\pi a_0 R^4 \exp(-\lambda t)}{\nu \alpha^2}\left[1 - \frac{2J_1(\alpha)}{\alpha J_0(\alpha)}\right]. \qquad (9)$$

In the limit of small Womersley numbers, from (9), it follows $Q = \pi a_0 R^4 \exp(-\lambda t)(1 - \frac{\alpha^2}{6} + O(\alpha^4))/8\nu$, that for $\lambda \to 0, \alpha \to 0$ exactly coincides with the well known dependency stated empirically by Hagen (G. Hagen, 1839) and Poiseuille (J.L.M. Poiseuille, 1840) and explained theoretically by Stokes (G.G. Stokes, 1845) [5]. It follows that in the considered limit nonstationarity of the fluid flow itself leads only to the increase of the hydro resistance $1/(1 - \frac{\alpha^2}{6} + ..)$ times. In the general case of arbitrary Womersley numbers, hydro resistance changes (decreases) $8(\frac{2J_1(\alpha)}{\alpha J_0(\alpha)} - 1)/\alpha^2$ times with respect to the case of zero Womersley number. For example, in the case of large Womersley numbers, when $\alpha \to \pi/2 + \pi n, n \gg 1$, hydro resistance tends to zero. More over, if conditions (5) or (6) are met, the value of this resistance becomes negative that leads to the necessity of qualitative change of the notion of hydro resistance for such modes of the fluid flow.

Specified above wide range of variability of the Womersley number $\alpha$ for the blood flow in CVS in [1] is estimated based on the assumption that for artery vessels of different caliber the value of $\lambda$ is one and the same and is defined only by the heart rate (HR). In the present work, vice versa, the value of the parameter $\lambda$ ( characterizing rate of monotonic exponential decay with time for the value of longitudinal pressure gradient) may in principle depend also on the radius of the corresponding vessel in the hierarchy of artery vessel system, not on HR only, coinciding with it by the order of value only for the largest arteries. In particular, we can suppose scale invariance of the Womersley number $\alpha$=const, under fixing which the value $\lambda \square O(\alpha^2 \nu / R^2)$ can significantly grow with the decrease of radius of the blood vessel R. For such case, in inequalities (5) or (6), values of n may be fixed and the inequalities themselves can have one and the same form for any vessel of artery vessel system independently on their radius.

Under conditions (5) or (6) direction of the impulse flow changes sign and it corresponds now to the transfer of impulse from the pipe wall to the flow. In the result, the flow can get additional acceleration but only during a finite time interval proportional to the value of $\lambda^{-1}$. Real wall of an artery vessel has not only passive elastic-resilient features but also its own muscular system regulated due to nervous and humoral factors. In the norm, it may provide longer support of the considered undamped blood flow mode thanks to the viscous friction forces and corresponding energy effective mode of CVS work. Really, in the case of the specified prolongation of the mode with negative friction resistance, it is possible its joining with the next heart cycle. Initial part of such a cycle (exhibiting before the start of the phase of monotonic decrease of longitudinal pressure gradient) can be modeled by more short time phase of sharp exponential monotonic increase of the value of longitudinal pressure gradient. Note that change of sign near $\lambda$ (i.e. for negative values of this parameter $\lambda < 0$) in (1) leads to the same structured solution of type (3), but in which it is necessary to replace $\lambda$ by $-\lambda$ and Bessel function $J_0$ by the modified Bessel function of zero order $I_0$. Meanwhile value (4) is negative for any value Womersley parameter $\alpha$.

Thus, only in the phase of monotonic decrease according to (1) and only under condition (6), it takes place the energy saving effect of realization of negative friction resistance leading to the transfer of impulse from the wall to the flow additionally accelerating it. In engineering practice similar energy saving effects for pulsating modes of the fluid flow (associated even with "super-fluidity") are known for a long time. They were used already in the development of energy optimal system of the fuel feeding for FAU missiles in 1941 and they are also used in the base of empirical equations of G. Poyedintsev-O. Voronova [7], used in the software of the device CARDIOCOD [8, 9] (certified in RF and states of EU). This device is used for complex cardio-metering (synchronized recording and analysis of ECG and rheography) for setting normal mode of functioning of CVS and for quantitative defining of deviations from the normal energy effective work of ECG.

In the result, conclusion about possibility of hydro-dynamic mechanism of sustaining in norm observed phenomenal efficiency of CVS providing by correlation of hydro-mechanical and nervous-humoral factors defining opportunity of the prolongation of the mode with negative friction resistance and its sequential joining in conjugated heart cycles. And, conversely, the problem of observed individual sensitivity to variable environmental factors gets new understanding on the base of the suggested mechanism that joins expressions of such sensitivity with particular individually defined reasons of disruptions of the pointed correlation in the work of CVS and other organism systems.

## 3. Stability of non-stationary hydro-dynamic mode

It is interesting to consider the question of stability of flow mode (3) to arbitrary small by amplitude disturbances on the base of respective generalization of the investigation of linear stability of the Hagen-Poiseuille (HP) flow. It is possible to use results obtained in [10,11], where it was suggested a new linear theory of hydrodynamic stability of the HP flow and the first time conditions of instability are obtained for very arbitrary small on amplitude axial-symmetrical disturbances already for above threshold Reynolds numbers Re>124. Contrary to the classical hydrodynamic stability theory [5], we consider not pure periodic along the pipe axis disturbances but quasi-periodic with two character incommensurable longitudinal periods each of which describes longitudinal periodic variability of one out of two independent radial Galerkin's modes of tangent velocity field disturbance.

Let us consider opportunity of generalization of that theory on the case of non-stationary modification of HP flow (3). We use (3) as the main disturbed flow instead of known (see [5]) classical exact stationary solution of Navie-Stokes equations ( $V_{0z} = V_{\max 0}(1 - \frac{r^2}{R^2}), V_{\max 0} = \frac{R^2}{4\rho_0 \nu} \frac{\partial p_0}{\partial z}$ ). Actually, in the limit of small $\alpha << 1$ from (3) it follows that $V_z = V_{0z} \exp(-\lambda t) \square V_{0z}(1 - \alpha^2 \tau + ...), \tau = t\nu / R^2$.

In cylindrical reference frame (z, r, $\varphi$) the equation for extremely small on amplitude disturbance of tangent velocity field component (in axially symmetrical case, i.e. when $\frac{\partial V_\varphi}{\partial \varphi} = 0$) has the form [10,11]:

$$\frac{\partial V_\varphi}{\partial t} + V_z(r,t)\frac{\partial V_\varphi}{\partial z} = \nu(\Delta V_\varphi - \frac{V_\varphi}{r^2}),  \qquad (10)$$

where $V_z$ is the main flow from (3), which is affected by the disturbance $V_\varphi$ so that the resultant flow is already spiral.

Let us find a solution of the equation (10), which satisfies the boundary non-slipping condition on the solid boundary of the pipe (i.e., the condition $V_\varphi(r = R, z, t) = 0$ must hold). The solution is represented in the following form:

$$V_\varphi = V_m \sum_{n=1}^{N} A_n(z,t) J_1(\gamma_{1,n}\frac{r}{R}), V_m = \frac{a_0}{\lambda}, \qquad (11)$$

where $J_1$ is the Bessel function of the first order, and $\gamma_{1,n}$ are its zeroes, i.e.. $J_1(\gamma_{1,n}) = 0$. After substitution of (11) in (10), multiplication by $rJ_1(\gamma_{1,m}\frac{r}{R})$ and integration over r in the limits from 0 to R, we get (in the result of application of this standard procedure of Galerkin's approximation of the solution of the equation (10)) in dimensionless form a closed system of equations for N unknown radial amplitudes $A_m, m = 1...N$:

$$\frac{\partial A_m}{\partial \tau} + \gamma_{1,m}^2 A_m - \frac{\partial^2 A_m}{\partial x^2} + \frac{4}{\alpha^2}\text{Re}\exp(-\alpha^2\tau)\sum_{n=1}^{N} p_{nm}\frac{\partial A_n}{\partial x} = 0,$$

$$p_{nm} = -\delta_{nm} + \frac{2}{J_2^2(\gamma_{1,m})}\int_0^1 dy\, y \frac{J_0(\alpha y)}{J_0(\alpha)} J_1(\gamma_{1,n} y) J_1(\gamma_{1,m} y). \qquad (12)$$

In (12) $\text{Re} = V_{\max} R / \nu, (V_{\max} \equiv \frac{a_0 R^2}{4\nu})$ is Reynolds number, $J_2$ is the Bessel function of the second order, $\delta_{nm}$ is the Kronecker symbol, x=z/R, $\tau = t\nu/R^2$. The system (12) exactly coincides with the system (3) in [11] when Womersley number tends to zero or equals zero $\alpha = 0$, when in (12) there must be made substitution $\frac{J_0(\alpha y)}{J_0(\alpha)} \to (1 + \alpha^2(1 - y^2)/4)$, and also $V_m$ must be replaced by $V_{\max 0}$ in the definition of Reynolds number Re.

Thus, in the limit of small Womersley numbers $\alpha \square 10^{-3}$ character for arteries of small caliber, conclusions obtained in [10,11] about linear instability leading to spiralization of the resultant flow for Re>124, actually are remained true also for the non-stationary main flow (3), considered here instead of the stationary mode of the HP flow. At the same time, for aorta, where the value of Womersley number $\alpha \square 10$, conclusions [10, 11] already need substantial refinement since in (12) the term proportional to the Reynolds number Re is found extremely small even for short times $\tau \leq 10^{-1}$. Obviously realization of the mode of linear exponential instability is found out to be difficult for the flow (3), modeling blood flow in aorta and other large vessels of the arterial system. However when considering time scales including not one but many heart cycles (when the pressure gradient in (1) is described by a periodic function of time, and using substitution $\lambda \to i\omega$ in (1)), there appear already qualitatively new opportunities for realization of instability of the main flow related with the effect of parametric resonant instability [12]. In that case it might be reasonable to use Krylov-Bogolyubov method of averaging in the region of the main demultiplicative resonance of frequencies (see [12]). More detailed investigation of stability of the flow (3) might be conducted in a separate work but already now there are grounds state that solution of the problem of stability of the flow (3) with respect to small disturbances and possibility of respective transformation of (3) in a flow with finite spirality substantially depends on the Womersley number.

## 4. Conclusions

Results obtained in the present paper point on important features of unsteady non-stationary modes of viscous incompressible fluid flow in the pipe with circular cross-section and solid walls. We show that in principle energy effective hydro-dynamic modes are possible that practically do not have losses due to friction on pipe walls under defined above conditions on Womersley number for such a flow. We obtained conditions of realization of stable to turbulent pulsations spiral swirled flow in the pipe emerging already for not large Reynolds numbers in the result of linear hydro-dynamic instability of the main non-stationary flow generalizing well known mode of the HP flow for the case of unsteady flow. In its turn, relative (compared to a pure laminar flow) energy efficiency of the swirled structural flow organization is caused by minimality of the rate of viscous dissipation of kinetic energy and by extremum of this flow energy [13, 14]. Actually, swirled spiral blood flow structures are observed in CVS and are found to be important for sustaining of it its normal function efficiency [14, 15]. The obtained conclusion on possibility of blood stream spiralization in arteries of small caliber (due to the linear hydro-dynamic instability of the flow (3)) complements existing now concept (see [15]) about realization of the swirled organization mainly in the magistral departments of CVS (left heart ventricle, aorta and arteries of the largest caliber).